\documentclass{article}
\usepackage{romp}
\usepackage{graphicx}
\usepackage{subfigure}
\usepackage{fancyheadings}
\usepackage{amsmath,dsfont}
\usepackage{amssymb}
\usepackage{cite}

\title{ Breakdown of separability due to confinement }
\author{ V.I. Man'ko \\ P.N. Lebedev Physical Institute, Russian Academy of Sciences\\
Leninskii Prospect 53, Moscow 119991, Russia\\ Moscow Institute of Physics and Technology\\
Institutskii Per. 9, Dolgoprudny Moscow Region 141700, Russia\\
e-mail:  manko@na.infn.it\\[2ex]
         L.A. Markovich \thanks{The study in section \ref{sec:3} and \ref{sec:4} by Markovich L.A. was supported by the Russian Science Foundation grant (14-50-00150).
}
                      \\
Institute for information transmission problems, Moscow\\
Bolshoy Karetny per. 19, build.1, Moscow 127051, Russia\\
Institute of Control Sciences, Russian Academy of Sciences\\
Profsoyuznaya 65, Moscow 117997, Russia \\ e-mail: kimo1@mail.ru \\[2ex]
        A. Messina \\
        Dipartimento di Fisica e Chimica, Universit\`{a} di Palermo\\
I-90123 Palermo, Italy\\
Sezione I.N.F.N., Catania \\ antonino.messina1949@gmail.com            }

\begin{document}

\maketitle
\begin{abstract}
A simple system of two particles in a bidimensional configurational space $S$ is studied. The possibility of breaking in $S$ the time independent Schr\"{o}dinger equation
of the system into two separated one-dimensional one-body Schr\"{o}dinger equations is assumed. In this paper, we focus on how the latter property is countered by imposing such boundary conditions as  confinement in a limited region of $S$ and/or restrictions on the joint coordinate probability density stemming from the sign-invariance  condition of the relative coordinate (an impenetrability condition). Our investigation demonstrates the reducibility of the problem under scrutiny into that of a single particle living in a limited domain of its bidimensional configurational space. These general ideas are illustrated introducing the  coordinates $X_c$ and $x$ of the center of mass of two particles and of the associated relative motion, respectively. The effects of the confinement and the impenetrability are then analyzed by studying with the help of an appropriate Green's function and the time evolution of the covariance of $X_c$ and $x$. Moreover, to calculate the state of the   single particle constrained within a square, a rhombus, a triangle and a rectangle the  Green's function expression in terms of Jacobi  $\theta_3$-function is applied. All the results are illustrated by examples.
\end{abstract}


\section{Introduction}
\par Over the decade, many papers appeared dedicated to the problem of quantum systems with boundaries. Several works are devoted to the problem of particles confined in a box, sometimes with moving walls \cite{Doescher,Pinder,Schlitt,Dodonov,Martino,Martino2,DodonovNikonov} or specific shapes \cite{Mousavi1,Mousavi2}. The confinement means the restriction on the motion of randomly moving particles, e.g. by the potential barriers.
\par The confined systems have become a topical issue common to many research areas from condensed matter physics and  quantum optics to biophysics.
The first easy to understand reason is that today an accurate quantitative prediction of the physical behavior of such  systems is required by   experimentalists since they cannot ignore confinement effects when the spacial  constraints stemming from the extreme miniaturization required for applications reach  micro or nano-sizes. During the recent years some laboratory realization of new experimental systems like two and three dimensional graphite cones, carbon nano-tube rings \cite{Guniberti,Zhang} and torus shaped nano-rings \cite{Garcia,Lorke} generated interest in the development of the idea of the quantum confinement. They can be used in development of nano and molecular electronic circuit devices.
The second important reason is that the  physical behavior of exemplary systems like a harmonic oscillator \cite{Sen,Gueorguiev,Amore}, an atom \cite{Cooper,Aquino,Fernandez} or a small molecule, when subjected to such boundary conditions, may exhibit  qualitative differences  with respect to  that available in the literature,   for example, due to a breaking of the geometrical symmetry. This  may lead to deep modifications on the eigensolutions of the system as well as to  the need of a different way of treating the center of mass motion in the case of more than one particle in the system.
\par In this paper, we first elucidate the breakdown of the separability of the center of mass motion from their relative motion for an unidimensional system of two, even non interacting, particles stemming from the confinement of the system. To this end, we first analyze the "traditional" confinement constraint consisting on limiting the motion of the two particles inside a finite and the same interval $I \subset \mathds{R}$. In addition to such a boundary condition we distinguish the case when the relative coordinate is allowed to assume both positive and negative values from that when one of two particles is always on the same side with respect to the other one. We refer to this last situation speaking of an "impenetrability condition" describing it as an effective further constraint on the system. Recently unidimensional systems exhibiting this kind of restriction on the motion of the particles have been realized in laboratory \cite{Dehkharghani}.  Our target is to compare the separable motion of two totaly unconstrained particles, that is $I= \mathds{R}$ and the lack of impenetrability, with the ones when at least one out of the two constraints  is instead present.
Our results clearly illustrate   that the existence of a basis of factorized stationary states of two even noninteracting quantum particles critically depends on whether and how the relevant dynamical variables get algebraically linked on the frontier of the bidimensional domain out of which any wave function compatible with the assumed constraints, vanish.
In other words, we highlight that separability depends not only on the structure of the relative Schr\"{o}dinger equation but also on the geometric shape of the normalization domain.
This observation paves the way to the possibility of tracing back the confined motion of our unidimensional system of two noninteracting particles to that of a fictitious particle moving in a plane within a domain whose shape is determined by the restrictions imposed on our original two-particle system. The second part of this paper is thus dedicated to quantum billiards problems with plain domains as the square, the rhombus, the triangle and the rectangle.
Also, the time evolution of the covariance of the center of mass coordinates is obtained. To this end, the theta-three Green's function theory \cite{Dodonov2} is applied. It is worthy to mention that Jacobi $\theta_3$-function was used to study coherence states of a charge moving in constant magnetic field \cite{Malkin}.
\par A quantum particle inside the potential of a special form is an interesting problem. There are many studies connected with the quantum and the classical properties of the
bidimensional difficult geometries like, for example, the  Robnik's billiard and the Bunimovich's stadium \cite{Robnik,Bunimovich}, the Sinai's billiard \cite{Sinai}. The quantum particle in the triangle potential is studied in \cite{Heller,Berry,Artuso,Casati}. The triangular shaped potential appears in different contexts. For example, in the equipotential curves for the H\'{e}non-Heiles system \cite{Aguirre}. Inside the boundary the potential behaves like the bidimensional harmonic oscillator, but in the billiard case, the particle has a free motion inside the triangular shaped domain.
The brief review on the triangular billiard geometry problems is given, for example, in \cite{Joseph}.
\\In this paper, we study four different shapes of the potential. It is shown how using the known boundary conditions for the particle confined in the square box with the side $d$, the boundary conditions for the boxes forming the rhombus, the triangle and the rectangle are constructed. Moreover, the Green's function using the Jacobi $\theta_3$-function for all four cases is obtained.
\par The paper is organized as follows. In  Sec. \ref{sec:1} the case of two noninteracting particles confined in the unidimensional box is considered. The boundary conditions are written for the center of mass motion. The case of the presence of the impenetrability condition is discussed. In Sec. \ref{sec:2} the problem of the single free  particle motion on the plane in the domain of the motion being restricted by the unpenetrated  walls forming the square, the rhombus, the triangle (the triangle billiard) and the rectangle are studied.
In Sec. \ref{sec:3} the time evolution of the wave function is obtained. The time evolution of the covariance of the center of mass coordinates is shown for the bounded and the unbounded problems. In Sec. \ref{sec:4} the Green's function is used to find the time-dependent states for the covariance.  The results obtained are illustrated on the example of the single particle motion confined in the square impenetrable box  with the side $d$.
\section{Two particles confined in a one-dimensional box} \label{sec:1}


Let us consider a system of two particles with masses $m_1$ and $m_2$ confined in a one-dimensional box (tube) of a length $d$ delimited by two impenetrable walls. The coordinates of the first and the second particle are $0\leq x_1\leq d$  and $0\leq x_2\leq d$, respectively.
The Schr\"{o}dinger equation for the two noninteracting particles is the following
\begin{eqnarray}\label{9}-\frac{\hbar^2}{2}\left(\frac{1}{m_1}\frac{\partial^2}{\partial x_1^2}+\frac{1}{m_2}\frac{\partial^2}{\partial x_2^2}\right)\Psi(x_1,x_2)=E\Psi(x_1,x_2).
\end{eqnarray}
It is known that in general the solution of (\ref{9}) is the following
\begin{eqnarray}\label{4}\Psi_{free}(x_1,x_2)=(A\sin(k_1x_1)+B\cos(k_1x_1))(C\sin(k_2x_2)+D\cos(k_2x_2)),
\end{eqnarray}
where $k_i=\sqrt{\frac{2m_iE}{\hbar^2}}$, $i=1,2$. The four constants $A,B,C,D$ must be determined taking into account normalization and the presence of the boundaries. In our case
$\Psi_{free}(x_1,x_2)$  must vanish outside of the box and on the boundaries, that is
\begin{eqnarray}\Psi(x_1=0,x_2)=\Psi(x_1,x_2=0)=0,\quad \Psi(x_1=d,x_2)=\Psi(x_1,x_2=d)=0.\label{31}\end{eqnarray}
The solution (\ref{4}) respecting such boundary conditions (\ref{31}) can be hence  rewritten as
\begin{eqnarray}\label{5}\Psi_{conf}(x_1,x_2)=N\sin(k_1x_1)\sin(k_2x_2),
\end{eqnarray}
where $k_1=\pi n_1/d$, $k_2=\pi n_2/d$, $N$ is the appropriate normalization constant.
\par
Introducing the well known center of mass transformation
\begin{eqnarray}\label{3}x=x_1-x_2,\quad X_c=\frac{m_1x_1+m_2x_2}{M},\quad M=m_1+m_2,\quad \mu=\frac{m_1m_2}{M},
\end{eqnarray}
one can rewrite (\ref{9}) as
\begin{eqnarray}\label{6}-\frac{\hbar^2}{2}\left(\frac{1}{\mu}\frac{\partial^2}{\partial x^2}+\frac{1}{M}\frac{\partial^2}{\partial X_c^2}\right)\widetilde{\Psi}(X_c,x)=E\widetilde{\Psi}(X_c,x).
\end{eqnarray}
For the problem of free particles this approach leads to the possibility of finding eigensolution of (\ref{6}) in the form of a product, e.g. $\widetilde{\Psi}(X_c,x)=\widetilde{\Psi}_1(X_c)\widetilde{\Psi}_2(x)$.
\par The representation of (\ref{5}) in terms (\ref{3}) is the following
\begin{eqnarray}\label{10}\widetilde{\Psi}_{conf}(X_c,x)&=&
-\widetilde{N}e^{-iX_c(k_1+k_2)}e^{ix\frac{m_2k_1-m_1k_2}{M}}\\\nonumber
&\times&\left(e^{2iX_ck_1}e^{-2ix\frac{m_2k_1}{M}}-1\right)\left(e^{2iX_ck_2}e^{2ix\frac{m_1k_2}{M}}-1\right),
\end{eqnarray}
where $\widetilde{N}$ is a normalization constant.
\par
\begin{figure}[ht]
\begin{center}
\begin{minipage}[ht]{0.60\linewidth}
\includegraphics[width=1\linewidth]{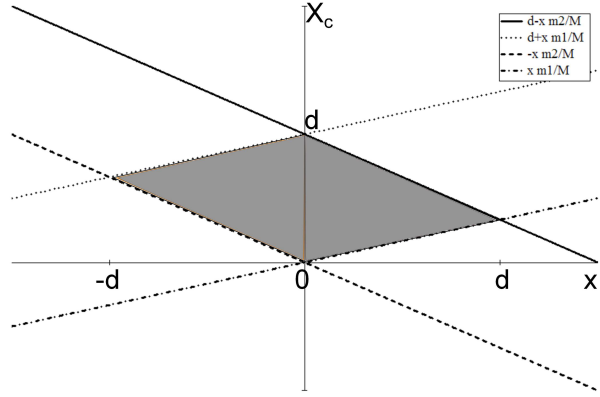}
\vspace{-4mm}
\caption{The domain of $X_c$ and $x$.}
\label{fig:2}
\end{minipage}
\end{center}
\end{figure}

 From (\ref{3}) we can write that
\begin{eqnarray}\label{25}x_1&=&X_c+\frac{m_2}{M}x,\quad x_2=X_c-\frac{m_1}{M}x,\quad 0\leq x_1\leq d,\quad 0\leq x_2\leq d.
\end{eqnarray}
Hence,  in the plane $(X_c, x)$ the domain of variability of $X_c$ and $x$ turn out to be bounded by the four lines shown in Fig.~\ref{fig:2}.
Moreover, we can deduce the geometrical boundary conditions for the solution of (\ref{6}) in the center of mass coordinates as
 \begin{eqnarray}\label{11}\widetilde{\Psi}\left(X_c,x|X_c+\frac{m_2}{M}x=d\right)=0,\quad \widetilde{\Psi}\left(X_c,x|X_c-\frac{m_1}{M}x=d\right)=0,\\
 \widetilde{\Psi}\left(X_c,x|X_c+\frac{m_2}{M}x=0\right)=0,\quad \widetilde{\Psi}\left(X_c,x|X_c-\frac{m_1}{M}x=0\right)=0\nonumber.
\end{eqnarray}
One should note that from the latter conditions the coordinates $X_c$ and $x$ are algebraically related in the frontier of the domain. Hence, in the presence of the boundary conditions the separation of the variables is impossible \cite{Tanner} and we can not look for the solution of (\ref{6}) as a factorized function of the two variables.
 Thus (\ref{6}) can not be rewritten in two differential equations and relative boundary conditions each one depending only on one variable. We conclude that in the presence of the boundary conditions (\ref{11}) the problem (\ref{6})  is not separable.
\par
\begin{figure}[ht]
\begin{center}
\begin{minipage}[ht]{0.60\linewidth}
\includegraphics[width=1\linewidth]{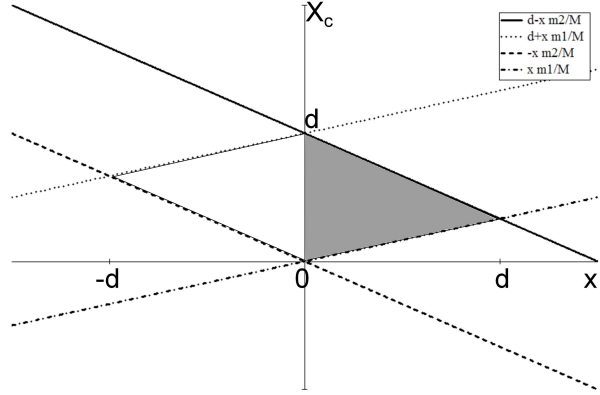}
\vspace{-4mm}
\caption{The domain of $X_c$ and $x$ when the particles can not penetrate each other.}
\label{fig:3}
\end{minipage}
\end{center}
\end{figure}

 Let us search the solution of (\ref{6}) in the form
\begin{eqnarray*}\widetilde{\Psi}(X_c,x)&=&(Ae^{\imath k1(aX_c+bx)}+Be^{-\imath k1(aX_c+bx)})(Ce^{\imath k2(cX_c+dx)}+De^{-\imath k2(cX_c+dx)}).
\end{eqnarray*}
Using the boundary conditions (\ref{11}), we get $a=c=1$, $A=C=1, B=D=-1$, $b=-m_2/M$, $d=m_1/M$ and $k_1=\pi n_1/d$, $k_2=\pi n_2/d$.
Hence, the energy eigenvalues are
\begin{eqnarray*}E_{n_1n_2}=\frac{\hbar^2 k_1^2}{2m_1}+\frac{\hbar^2 k_2^2}{2m_2}.\end{eqnarray*}
Thus, the solution of (\ref{6}) with respect to the new geometric boundary conditions (\ref{11}) is the following
\begin{eqnarray}\label{17}\widetilde{\Psi}_{n_1n_2}(X_c,x)\!\!&=&\!\!\widetilde{C}(e^{\imath \frac{\pi n_1}{d}(X_c+\frac{m_2}{M}x)}-e^{-\imath \frac{\pi n_1}{d}(X_c+\frac{m_2}{M}x)})(e^{\imath \frac{\pi n_2}{d}(X_c-\frac{m_1}{M}x)}-e^{-\imath \frac{\pi n_2}{d}(X_c-\frac{m_1}{M}x)})\nonumber\\
&=&\widetilde{A}\sin\left(\frac{\pi n_1}{d}(X_c+\frac{m_2}{M}x)\right)\sin\left(\frac{\pi n_2}{d}(X_c-\frac{m_1}{M}x)\right),
\end{eqnarray}
that is (\ref{5}) rewritten in the center of mass coordinates.
The normalization constant can be found from normalization condition, rewritten in the notations  (\ref{3}). Namely
\begin{eqnarray*}2\int\limits_{0}^{d}dx\left(\int\limits_{\frac{m_1x}{M}}^{d-\frac{m_2x}{M}}\Big|\widetilde{\Psi}_{n_1n_2}(X_c,x)\Big|^2dX_c\right)=1
\end{eqnarray*}
and the constant may be assumed to be equal to $\widetilde{A}=\frac{2}{d}$. 
If we assume that the crossection of the tubes possese a radius comparable with the liniar dimension of the particles, then they can not penetrate each other. Hence, we can think that the first particle is on the right hand side while the second one is on the left hand side of the box, e.g. $x_1>x_2$. Using this additional condition and (\ref{25}) we can write
\begin{eqnarray}X_c+\frac{m_2}{M}x>X_c-\frac{m_1}{M}x.\label{38}
\end{eqnarray}
Thus $x>0$ that is always true for the present system, where $0<x<d$. That means that the additional condition on the penetrability of the particles influences the boundary conditions for the center of mass motion and the new domain is shown in Fig.\ref{fig:3}. Note that (\ref{17}) does not fullfill the latter geometric boundary condition.
In the presence of the inpenetrability condition, e.g. $x_1>x_2$, but absence of the confinement (that means that $x_1,x_2\in(-\infty,\infty)$) we can write (\ref{38}) and get the condition analogues to confinement from the one boundary.
\section{One particle confined in the two dimensional box}\label{sec:2}
\begin{figure}[ht]
\begin{center}
\begin{minipage}[ht]{0.2\linewidth}
\includegraphics[width=1\linewidth]{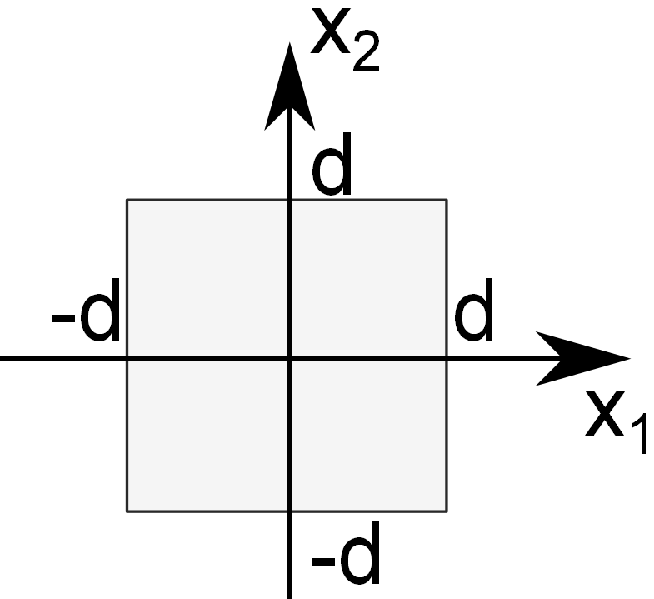}
\vspace{-4mm}
\end{minipage}
\hfill
\begin{minipage}[Ht]{0.2\linewidth}
\includegraphics[width=1\linewidth]{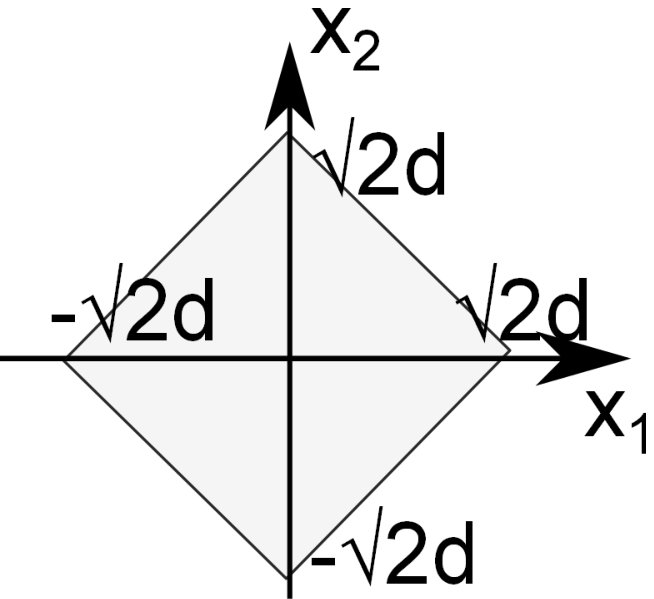}
\vspace{-4mm}
\end{minipage}
\hfill
\begin{minipage}[ht]{0.2\linewidth}
\includegraphics[width=1\linewidth]{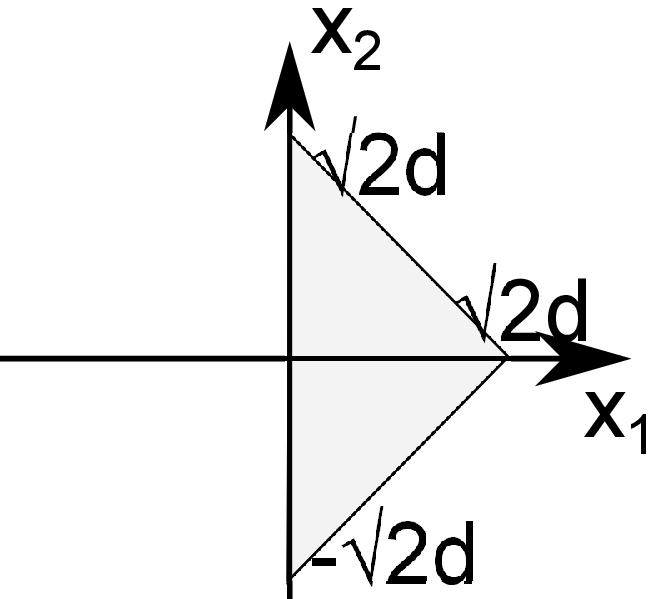}
\vspace{-4mm}
\end{minipage}
\hfill
\begin{minipage}[ht]{0.2\linewidth}
\includegraphics[width=1\linewidth]{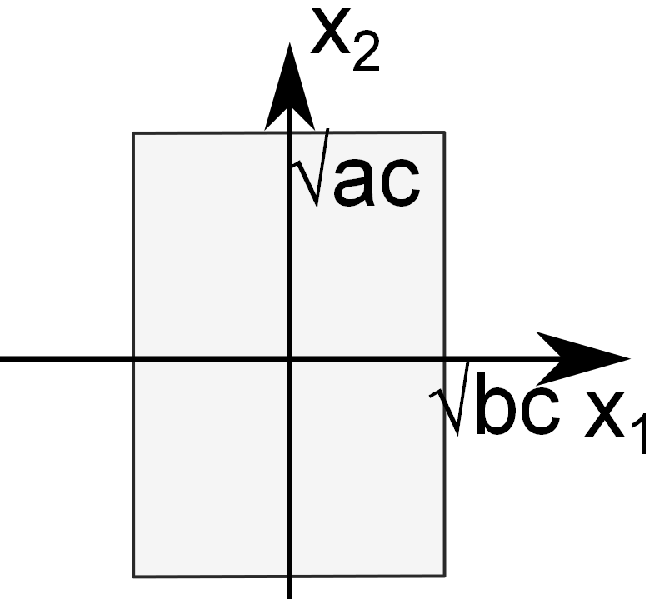}
\vspace{-4mm}
\end{minipage}
 \caption{The domains of $x_1,x_2$.\label{fig:46}}
\end{center}
\end{figure}

The problem of the two particles confined in the one dimensional box is equvivalent to the problem of one particle confined in the two dimensional box. The Schr\"{o}dinger equation for such system is
 \begin{eqnarray}-\frac{1}{2m}\left(\frac{\partial^2}{\partial x_1^2}+\frac{\partial^2}{\partial x_2^2}\right)\Phi(x_1,x_2)=E\Phi(x_1,x_2)\label{41}.
\end{eqnarray}
Hence, in this section we study the the single free  particle motion on the plane, but the domain of the motion being restricted by unpenetrated  walls forming the
the square with the side $d$, the rhombus with the side $d$, the triangle and the rectangle (see Fig. \ref{fig:46}).
\par Let us start from the square impenetrable box  with the side $d$. The solution of  the Schr\"{o}dinger equation \eqref{41} is
  \begin{eqnarray}\Phi(x_1,x_2)&=&A\sin(n_1x_1)\sin(n_2x_2),\label{45}\end{eqnarray}
  where $A$ is a normalization parameter.
  From the boundary conditions  \begin{eqnarray*}\Phi(x_1=\pm d,x_2)=\Phi(x_1,x_2=\pm d)=0\end{eqnarray*}
 we can estimate the parameters of the solution, ss. $n_1=\frac{\pi N_1}{d}$, $n_2=\frac{\pi N_2}{d}$, $N_1,N_2=0,\pm 1,\pm2,\ldots$.
\par  If we rotate the square box on $45^{\circ}$ we will get the rhombus with the side $d$.
As the solution of the Schr\"{o}dinger equation \eqref{41} for such system we can take the following function
  \begin{eqnarray}\widetilde{\Phi}(x_1,x_2)&=&B\sin(n_1(x_1+x_2))\sin(n_2(x_1-x_2)),\label{42}\end{eqnarray}
  where $B$ is a normalization parameter.
  The boundary conditions are   \begin{eqnarray*}\widetilde{\Phi}(x_1,x_2=\pm x_1+\sqrt{2}d)=\widetilde{\Phi}(x_1,x_2=\pm x_1-\sqrt{2}d)=0.\end{eqnarray*}
     Hence, the parameters of the solution \eqref{42} are $n_1=\frac{\pi N_1}{\sqrt{2}d}$, $n_2=\frac{\pi N_2}{\sqrt{2}d}$, $N_1,N_2=0,\pm 1,\pm2,\ldots$.
\par Moreover, dividing the rhombus domain at zero boundary, we get the triangle box. The latter case is called the quantum triangle billiard. As the initial step we study the billiard where the sides of the triangle  are obtained by the intersections of the three lines
 described by the following equations
\begin{eqnarray}x_1=0,\quad x_2=x_1-\sqrt{2}d,\quad x_2=-x_1+\sqrt{2}d\label{40}
\end{eqnarray}
Hence, the boundary conditions for such system are
   \begin{eqnarray*}f(x_1=0,x_2)=f(x_1,x_2=\pm x_1+\sqrt{2}d)&=&f(x_1,x_2=\pm x_1-\sqrt{2}d)=0.\end{eqnarray*}
One should note here that all the latter conditions must be fulfilled simultaneously, which means that the solutions are equal to zero on the sides of the triangle given by \eqref{40}. As the solution of the Schr\"{o}dinger equation \eqref{41} that satisfy the latter conditions let us select the following function
\begin{eqnarray}f(x_1,x_2)&=&K(\widetilde{\Phi}(x_1,x_2)-\widetilde{\Phi}(-x_1,x_2)),\label{43}\end{eqnarray}
where $K$ is a normalization parameter. The other parameters of the solution \eqref{43} are $n_1=\frac{\pi N_1}{\sqrt{2}d}$, $n_2=\frac{\pi N_2}{\sqrt{2}d}$, $N_1,N_2=0,\pm 1,\pm2,\ldots$.
\par Finnaly, let us introduce the boundary conditions for  the rectangle box. To this end we first study the square box with the side $d$ and the Schr\"{o}dinger equation of the following form
   \begin{eqnarray}-\left(\frac{1}{2a}\frac{\partial^2}{\partial x_1^2}+\frac{1}{2b}\frac{\partial^2}{\partial x_2^2}\right)\Phi(x_1,x_2)=E\Phi(x_1,x_2)\label{44}
\end{eqnarray}
with the parameters of the Hamiltonian $a$ and $b$. The energy is $E_{N_1,N_2}=\frac{1}{2}\left(\frac{n_1^2}{a}+\frac{n_2^2}{b}\right)$, $n_1=\frac{\pi N_1}{d}$, $n_2=\frac{\pi N_2}{d}$, $N_1,N_2=0,\pm 1,\pm2,\ldots$ and the solution of the latter equation is given by \eqref{45}. Let us make the transformation to the new variables
$y_1=\sqrt{a}x_1$, $y_2=\sqrt{b}x_2$. The transformed  Schr\"{o}dinger equation \eqref{44} is the folowing
   \begin{eqnarray*}-\frac{1}{2}\left(\frac{\partial^2}{\partial y_1^2}+\frac{\partial^2}{\partial y_2^2}\right)\Phi\left(\frac{y_1}{\sqrt{a}},\frac{y_2}{\sqrt{b}}\right)=E\Phi\left(\frac{y_1}{\sqrt{a}},\frac{y_2}{\sqrt{b}}\right),
\end{eqnarray*}
where the solution can be written as
   \begin{eqnarray*}\Phi\left(\frac{y_1}{\sqrt{a}},\frac{y_2}{\sqrt{b}}\right)&=&A\sin\left(n_1\frac{y_1}{\sqrt{a}}\right)\sin\left(n_2\frac{y_2}{\sqrt{b}}\right)\end{eqnarray*}
Thus, the boundary conditions for the rectangle domain with the sides $d\sqrt{a}$, $d\sqrt{b}$ are
   \begin{eqnarray*}\Phi\left(y_1=\pm d\sqrt{a},\frac{y_2}{\sqrt{b}}\right)&=&\Phi\left(\frac{y_1}{\sqrt{a}},y_2=\pm d\sqrt{b}\right)=0.\end{eqnarray*}
We can conclude, that one can obtained the boundary conditions for the specific forms of boxes using the knowledge about the boundary conditions of the well known squar box case.
\section{Time evolution of the center-of-mass covariance}\label{sec:3}
Since the center of mass coordinates are dependent we are interested in there covariance time evolution for the bounded and unbounded problems.
By definition, the square modulus of the wave function  is the probability density function
\begin{eqnarray*}f(X_c,x,t)&=&\Big|\Phi(X_c,x,t)\Big|^2.
\end{eqnarray*}
Thus, the expectations for $X_c$, $x$ and $X_c\cdot x$ are the following
\begin{eqnarray*}\mathbb{E}(x)&=&\int\limits_{-\infty}^{\infty} d\widetilde{x} \int\limits_{-\infty}^{\infty}\widetilde{X} f(\widetilde{X},x,t)d\widetilde{X},\quad
\mathbb{E}(X_c)=\int\limits_{-\infty}^{\infty} d\widetilde{x} \int\limits_{-\infty}^{\infty} \widetilde{x} f(X_c,\widetilde{x},t)d\widetilde{X},
\end{eqnarray*}
\begin{eqnarray*}
\mathbb{E}(X_c\cdot x)&=&\int\limits_{-\infty}^{\infty} d\widetilde{x} \int\limits_{-\infty}^{\infty}\widetilde{X}\widetilde{x} f(\widetilde{X},\widetilde{x},t)d\widetilde{X}
\end{eqnarray*}
and the covariance of $X_c$ and $x$ is
\begin{eqnarray}\label{36}Cov(X_c,x,t)&=&\mathbb{E}(X_c\cdot x)-\mathbb{E}(X_c)\mathbb{E}(x)
=\langle\Phi(X_c,x,t)|X_c x|\Phi(X_c,x,t)\rangle\\
&-&\langle\Phi(X_c,x,t)|X_c|\Phi(X_c,x,t)\rangle\langle\Phi(X_c,x,t)|x|\Phi(X_c,x,t)\rangle.\nonumber
\end{eqnarray}
However, we can just substitute the definition of $X_c$ and $x$ in \eqref{36} and get the equivalent result
\begin{eqnarray}\label{37}Cov(X_c,x,t)&=&\mathbb{E}((\mu_1x_1+\mu_2x_2)(x_1-x_2))-\mathbb{E}(\mu_1x_1+\mu_2x_2)\mathbb{E}(x_1-x_2)\\\nonumber
&=&\mu_1\left(\mathbb{E}(x_1^2)-\mathbb{E}^2(x_1)\right)-\mu_2\left(\mathbb{E}(x_2^2)-\mathbb{E}^2(x_2)\right)\\\nonumber
&=&\mu_1\left(\langle\Phi(x_1,x_2,t)|x_1^2|\Phi(x_1,x_2,t)\rangle-\left(\langle\Phi(x_1,x_2,t)|x_1|\Phi(x_1,x_2,t)\rangle\right)^2\right)\\\nonumber
&-&\mu_2\left(\langle\Phi(x_1,x_2,t)|x_2^2|\Phi(x_1,x_2,t)\rangle-\left(\langle\Phi(x_1,x_2,t)|x_2|\Phi(X_c,x,t)\rangle\right)^2\right).\nonumber
\end{eqnarray}
In the case of the equal masses the covariance is zero.
Note that we must change the variables in the state function before integrating \eqref{37}.
\subsection{Examples}
To obtain the covariance, one need to select some time state $\Phi(X_c,x,t)$ ($\Phi(X_c,x,t)$). Using the evolution operator we can obtain any state $|\Phi(t)\rangle$ from the initial one  $|\Phi(0)\rangle$ as
\begin{eqnarray*}|\Phi(X_c,x,t)\rangle=e^{-\frac{iHt}{\hbar}}|\Phi(X_c,x,0)\rangle.\end{eqnarray*}
 The evolution operator can be written as
\begin{eqnarray*}e^{-\frac{iHt}{\hbar}}&=&
e^{-\frac{i}{\hbar}\frac{P^2}{2M}t}e^{-\frac{i}{\hbar}\frac{p^2}{2\mu}t}.
\end{eqnarray*}
If the problem is unbounded the well known Baker-Hausdorff formula \cite{Miller} can be used $e^{\xi\widehat{A}}\widehat{B}e^{-\xi\widehat{A}}=\widehat{B}+\xi[\widehat{A}\widehat{B}]+\frac{\xi^2}{2}[\widehat{A}[\widehat{A}\widehat{B}]]+\ldots$
and
\begin{eqnarray}\label{28}e^{\frac{iHt}{\hbar}}X_cxe^{-\frac{iHt}{\hbar}}=\left(e^{\frac{i}{\hbar}\frac{P^2}{2M}t}X_ce^{-\frac{i}{\hbar}\frac{P^2}{2M}t}\right)\left(e^{\frac{i}{\hbar}\frac{p^2}{2\mu}t}xe^{-\frac{i}{\hbar}\frac{p^2}{2\mu}t}\right)
=\left(X_c+\frac{P}{M}t\right)\left(x+\frac{p}{\mu}t\right)
\end{eqnarray}
holds.
Hence, the expectations  are
\begin{eqnarray*}\langle\Phi(t)|X_c x|\Phi(t)\rangle&=&\langle\Phi(0)|e^{\frac{iHt}{\hbar}}X_cxe^{-\frac{iHt}{\hbar}}|\Phi(0)\rangle=
\langle\Phi(0)|\left(X_c+\frac{P}{M}t\right)\left(x+\frac{p}{\mu}t\right)|\Phi(0)\rangle,\\
\langle\Phi(t)|X_c|\Phi(t)\rangle&=&\langle\Phi(0)|e^{\frac{iHt}{\hbar}}X_ce^{-\frac{iHt}{\hbar}}|\Phi(0)\rangle=
\langle\Phi(0)|\left(X_c+\frac{P}{M}t\right)|\Phi(0)\rangle,\\
\langle\Phi(t)|x|\Phi(t)\rangle&=&\langle\Phi(0)|e^{\frac{iHt}{\hbar}}xe^{-\frac{iHt}{\hbar}}|\Phi(0)\rangle=
\langle\Phi(0)|\left(x+\frac{p}{\mu}t\right)|\Phi(0)\rangle
\end{eqnarray*}
and the covariance of $X_c$ and $x$ for the unbounded problem evolve in time as
\begin{eqnarray}\label{20}Cov(X_c,x,t)&=&
\langle\Phi(0)|\left(X_c+\frac{P}{M}t\right)\left(x+\frac{p}{\mu}t\right)|\Phi(0)\rangle\\\nonumber
&-&\langle\Phi(0)|\left(X_c+\frac{P}{M}t\right)|\Phi(0)\rangle
\langle\Phi(0)|\left(x+\frac{p}{\mu}t\right)|\Phi(0)\rangle.
\end{eqnarray}
\par For the bounded problem we can not use the Baker-Hausdorff formula since the  variables $X_c$ and $x$ are dependent.
As an example, let us take as the state $\Phi(X_c,x,t)$ the function that depends only from $\{n_1,n_2\}=\{(1,1),(2,2)\}$ and has the following view
\begin{eqnarray*}
\Phi(X_c,x,t)&=&A_1\widetilde{\Psi}_{11}(X_c,x)e^{-iE_{11}t/\hbar}+A_2\widetilde{\Psi}_{22}(X_c,x)e^{-iE_{22}t/\hbar}.
\end{eqnarray*}
Hence, the probability density function is
\begin{eqnarray*}
f(X_c,x,t)&=&A_1^2\widetilde{\Psi}^2_{11}(X_c,x)+A_2^2\widetilde{\Psi}^2_{22}(X_c,x)\\
&\cdot&+2A_1A_2\widetilde{\Psi}_{11}(X_c,x)\widetilde{\Psi}_{22}(X_c,x)\cos\left((E_{11}-E_{22})t/\hbar\right).
\end{eqnarray*}
From the normalization condition we can choose $A_1=A_2=\frac{2}{d}$.
Since for the bounded problem the coordinates $X_c$  and $x$ are in the domain shown in Fig. \ref{fig:2}, the covariance \eqref{36} is the following
\begin{eqnarray*}
Cov(X_c,x,t)&=&-\frac{d^2(m_1 - m_2)}{165888\pi^4M}\Bigg((668288- 61440\pi^2)\cos\left((E_{11}-E_{22})t/\hbar\right) \\
&+& 102400\cos^2\left((E_{11}-E_{22})t/\hbar\right) + 8640\pi^2 - 4608\pi^4 + 50625\Bigg),
\end{eqnarray*}
where $E_{11}-E_{22}=\frac{-3\hbar^2\pi^2}{2d^2\mu}$, $\mu=m_1m_2/M$. Certainly, the latter example is given for the  illustrative purposes. To get the more general covariance
we need to find the state $|\Phi(t)\rangle$ for the bounded problem. To this end, the Green's function theory can be applied.
\section{Green's function and time evolution}\label{sec:4}
In \cite{Fulling} the propagator is investigated for the case of the one particle confined in the box.
The relation between the wave function at time $t$ and $0$ is given by the following formula
\begin{eqnarray*}&&\Psi(X_c,x,t)=i\hbar\int\limits_{-\infty}^{\infty}\Psi(x_1,x_2)G(X_c,x,x_1,x_2,t)dx_1dx_2,
\end{eqnarray*}
where $G(X_c,x,x_1,x_2,t)$ is a time dependent Green's function or the propagator. In \cite{Hannesson} the Green's function is represented in terms of theta-three function
\begin{eqnarray*}G(\xi,\eta,\tau)&=&\frac{1}{2d}(\theta_3(\xi,\tau)-\theta_3(\eta,\tau)),
\end{eqnarray*}
where the theta-three function is
\begin{eqnarray*}\theta_3(\zeta,\tau)&=&\sum\limits_{n=-\infty}^{\infty}e^{2i n\zeta}e^{i \pi n^2\tau}.
\end{eqnarray*}
In the Whittaker and Watsou notation \cite{Whittaker} it can be represented as
\begin{eqnarray}\label{46}\theta_3(\zeta,q)&=&1+2\sum\limits_{n=1}^{\infty}\cos(2n\zeta)q^{n^2},\quad q=e^{i \pi\tau}.
\end{eqnarray}
Hence, the propagator for the one particle is
\begin{eqnarray*}G(x,x',t)&=&\frac{1}{2d}\sum\limits_{n=-\infty}^{\infty}e^{2i n(x-x')}e^{i \pi n^2t}.
\end{eqnarray*}
It is obvious, that the propagator for the two particles can be written as
\begin{eqnarray*}&&G(X_c,x,x_1,x_2,t)=\frac{1}{4d^2}(\theta_3(X_c,t)-\theta_3(x_1,t))(\theta_3(x,t)-\theta_3(x_2,t))\\
&=&\frac{1}{4d^2}\left(\sum\limits_{n=-\infty}^{\infty}e^{2i nX_c}e^{i \pi n^2t}-\sum\limits_{n=-\infty}^{\infty}e^{2i nx_1}e^{i \pi n^2t}\right)\\
&\cdot&\left(\sum\limits_{k=-\infty}^{\infty}e^{2i kx}e^{i \pi k^2t}-\sum\limits_{k=-\infty}^{\infty}e^{2i kx_2}e^{i \pi k^2t}\right).
\end{eqnarray*}
As the initial state $\Psi(x_1,x_2,0)$ one can select the following function
\begin{eqnarray*}\Phi(x_1,x_2,0)&=&\pm\frac{1}{\sqrt{2}d}\left(\sin\left(\frac{\pi}{d}x_1\right)\sin\left(\frac{\pi}{d}x_2\right)
+\sin\left(\frac{2\pi}{d}x_1\right)\sin\left(\frac{2\pi}{d}x_2\right)\right).
\end{eqnarray*}
Thus, we can write
\begin{eqnarray*}&&\Psi(X_c,x,t)=\frac{i\hbar}{4d^2}\Bigg(\theta_3(X_c,t)\theta_3(x,t)\int\limits_{0}^{d}\int\limits_{0}^{d}\Psi(x_1,x_2,0)dx_1dx_2\\
&-&\theta_3(x,t)\int\limits_{0}^{d}\int\limits_{0}^{d}\Psi(x_1,x_2,0)\theta_3(x_1,t)dx_1dx_2\\
&\cdot&-\theta_3(X_c,t)\int\limits_{0}^{d}\int\limits_{0}^{d}\Psi(x_1,x_2,0)
\theta_3(x_2,t)dx_1dx_2\\
&+&\int\limits_{0}^{d}\int\limits_{0}^{d}\Psi(x_1,x_2,0)\theta_3(x_1,t)\theta_3(x_2,t)dx_1dx_2\Bigg)\\
&=&\frac{i\hbar\sqrt{2}}{4d}\Bigg(\theta_3(X_c,t)\theta_3(x,t)\frac{2}{\pi^2}-(\theta_3(X_c,t)+\theta_3(x,t))\sum\limits_{n=-\infty}^{\infty}\frac{1+e^{2idn}}{\pi^2-4d^2n^2}e^{i \pi n^2t}\\
&+&\frac{\pi^2}{2}\left(\sum\limits_{n=-\infty}^{\infty}\frac{e^{2idn}+1}{\pi^2-4d^2n^2}e^{i \pi n^2t}\right)^2
+\frac{\pi^2}{8}\left(\sum\limits_{n=-\infty}^{\infty}\frac{e^{2idn}-1}{\pi^2-d^2n^2}e^{i \pi n^2t}\right)^2\Bigg).
\end{eqnarray*}
Using that \begin{eqnarray*}&&\frac{1}{ix}=\int_{0}^{\infty}\exp\left(-izx\right)dz\end{eqnarray*} we can write \begin{eqnarray*}&&\frac{1}{i(\pi^2-4d^2n^2)}=\int_{0}^{\infty}\exp\left(-iz(\pi^2-4d^2n^2)\right)dz.\end{eqnarray*}
Substituting $\Psi(X_c,x,t)$ in \eqref{36} the covariance time evolution can be obtained.
\subsection{The Green's function and the special boundary shapes}
Let us find the Green's function for the four boundary shapes presented in Sec. \ref{sec:2}
By definition the Green's function is the following
\begin{eqnarray}\label{39}G(x_1,x_2,x_1',x_2',t)=\sum\limits_{N_1=-\infty}^{\infty}\sum\limits_{N_2=-\infty}^{\infty}\Phi_{N_1,N_2}(x_1,x_2,t)\Phi^{\ast}_{N_1,N_2}(x_1',x_2')\exp(-iE_{N_1,N_2}t).
\end{eqnarray}
For the square box domain the energy is $E_{N_1,N_2}=\frac{n_1^2+n_2^2}{2m}=\frac{\pi^2(N_1^2+N_2^2)}{2md^2}$, $n_1=\frac{\pi N_1}{d}$, $n_2=\frac{\pi N_2}{d}$, $N_1,N_2=0,\pm 1,\pm2,\ldots$. Hence, for the latter boundary shape the Green's function \eqref{39} is the following
\begin{eqnarray*}&&G(x_1,x_2,x_1',x_2',t)=A^2\sum\limits_{N_1,N_2=-\infty}^{\infty}\sin (n_1x_1)\sin(n_2x_2)\sin(n_1x_1')
\sin(n_2x_2')e^{-iE_{N_1N_2t}}\\
&=&\frac{A^2}{4}\sum\limits_{N_1=-\infty}^{\infty}\left(\cos\left(\frac{\pi N_1}{d}(x_1+x_1')\right)
-\cos\left(\frac{\pi N_1}{d}(x_1-x_1')\right)\right)e^{-\frac{i\pi^2 N_1^2}{2md^2} t}\\
&\cdot&\sum\limits_{N_2=-\infty}^{\infty}\left(\cos\left(\frac{\pi N_2}{d}(x_2+x_2')\right)
-\cos\left(\frac{\pi N_2}{d}(x_2-x_2')\right)\right)e^{-\frac{i\pi^2 N_2^2}{2md^2} t}
\end{eqnarray*}
where we used the known formula $\sin\alpha\sin\beta=\frac{1}{2}\left(\cos(\alpha-\beta)-\cos(\alpha+\beta)\right)$.
Using the following notations
\begin{eqnarray*}&&\eta_1=\frac{\pi}{2d}(x_1-x_1'),\quad \eta_2=\frac{\pi}{2d}(x_1+x_1'),
\quad \eta_3=\frac{\pi}{2d}(x_2-x_2'),\\
&& \eta_4=\frac{\pi}{2d}(x_2+x_2'),\quad \tau=\frac{-\pi t}{2md^2}.
\end{eqnarray*}
we can rewrite the latter Green's function in a short form
\begin{eqnarray*}G(\eta_1,\eta_2,\eta_3,\eta_4,\tau)&=&\frac{A^2}{4}\sum\limits_{N_1=-\infty}^{\infty}\left(\cos(2\eta_1N_1)
-\cos(2\eta_2N_1)\right)e^{i\pi \tau N_1^2}\\
&\cdot&\sum\limits_{N_2=-\infty}^{\infty}\left(\cos(2\eta_3N_2)
-\cos(2 \eta_4N_2)\right)e^{i\pi \tau N_2^2}.
\end{eqnarray*}
Since $N_1=0,\pm 1,\pm2,\ldots$ we can split the latter sums into two sums as
 \begin{eqnarray*}G(\eta_1,\eta_2,\eta_3,\eta_4,\tau)&=&\frac{A^2}{4}\Bigg(\sum\limits_{N_1=1,2,\ldots}^{\infty}\left(\cos(2\eta_1N_1)
-\cos(2 \eta_2N_1)\right)e^{i\pi \tau N_1^2}\\
&+&\sum\limits_{N_1=-1,-2,\ldots}^{-\infty}\left(\cos(2\eta_1N_1)
-\cos(2 \eta_2N_1)\right)e^{i\pi \tau N_1^2}\Bigg)\\
&\cdot&\Bigg(
\sum\limits_{N_2=1,2,\ldots}^{\infty}\left(\cos(2 \eta_3N_2)
-\cos(2\eta_4N_2)\right)e^{i\pi \tau N_2^2}\\
&+&\sum\limits_{N_2=-1,-2,\ldots}^{\infty}\left(\cos(2\eta_3N_2)
-\cos(2\eta_4N_2)\right)e^{i\pi \tau N_2^2}\Bigg).
\end{eqnarray*}
Using the definition of the theta three function \eqref{46} and the notation $q=\exp(i\pi\tau)$ we can finally write that the Green's function for the square box with the side $d$ is
  \begin{eqnarray*}G(\eta_1,\eta_2,\eta_3,\eta_4,q)&=&\frac{A^2}{4}\left(\theta_3(\eta_1,q)-\theta_3(\eta_2,q)\right)
  \left(\theta_3(\eta_3,q)-\theta_3(\eta_4,q)\right).
\end{eqnarray*}
\par For the rhombus box we can  use the similar technics as in the previous case. The energy is $E_{N_1,N_2}=\frac{n_1^2+n_2^2}{m}=\frac{\pi^2(N_1^2+N_2^2)}{2md^2}$, $n_1=\frac{\pi N_1}{\sqrt{2}d}$, $n_2=\frac{\pi N_2}{\sqrt{2}d}$, $N_1,N_2=0,\pm 1,\pm2,\ldots$ and the Green's function for the rhombus with the side $d$ is the following
   \begin{eqnarray*}G_r(\eta_1,\eta_2,\eta_3,\eta_4,q)&=&B^2\left(\theta_3(\zeta_1,q)-\theta_3(\zeta_2,q)\right)
  \left(\theta_3(\zeta_3,q)-\theta_3(\zeta_4,q)\right),
\end{eqnarray*}
where we used the notations
\begin{eqnarray*}&&\zeta_1=\frac{\pi}{2\sqrt{2}d}(x_1-x_2-x_1'-x_2'),\quad \zeta_2=\frac{\pi}{2\sqrt{2}d}(x_1+x_2+x_1'+x_2'),\quad \tau=\frac{-\pi t}{2md^2}\\
&& \zeta_3=\frac{\pi}{2\sqrt{2}d}(x_1-x_2-x_1'+x_2'),\quad \zeta_4=\frac{\pi}{2\sqrt{2}d}(x_1-x_2+x_1'-x_2'),\quad q=e^{i\pi\tau}.
\end{eqnarray*}
\par For the triangle billiard the energy is equal to $E_{N_1,N_2}=2\frac{n_1^2+n_2^2}{m}=\frac{\pi^2(N_1^2+N_2^2)}{md^2}$, $n_1=\frac{\pi N_1}{\sqrt{2}d}$, $n_2=\frac{\pi N_2}{\sqrt{2}d}$, $N_1,N_2=0,\pm 1,\pm2,\ldots$, $q=\exp(2i\pi\tau)$ and the corresponding Green's function
      \begin{eqnarray*}G_b(\eta_1,\eta_2,\eta_3,\eta_4,q)&=&K^2\Bigg(\left(\theta_3(\zeta_1,q)-\theta_3(\zeta_2,q)\right)
  \left(\theta_3(\zeta_3,q)-\theta_3(\zeta_4,q)\right)\\
  &-&\left(\theta_3(-\zeta_1,q)-\theta_3(-\zeta_2,q)\right)
  \left(\theta_3(-\zeta_3,q)-\theta_3(-\zeta_4,q)\right)\Bigg)
\end{eqnarray*}
holds.
\par Finally, it is easy to verify, that for the rectangle domain the Green's function is
   \begin{eqnarray*}G_b(\xi_1,\xi_2,\xi_3,\xi_4,q_a,q_b)&=&
   A^2\left(\theta_3(\xi_1,q_a)-\theta_3(\xi_2,q_a)\right)
  \left(\theta_3(\xi_3,q_b)-\theta_3(\xi_4,q_b)\right),
\end{eqnarray*}
where the following notations were introduced
\begin{eqnarray*}&&\xi_1=\frac{\pi}{2\sqrt{a}d}(y_1-y_1'),
\quad \xi_2=\frac{\pi}{2\sqrt{a}d}(y_1+y_1'),\quad \tau_a=\frac{-\pi t}{2ad^2},\quad \tau_b=\frac{-\pi t}{2bd^2}\\
&& \xi_3=\frac{\pi}{2\sqrt{a}d}(y_2-y_2'),
\quad \xi_4=\frac{\pi}{2\sqrt{a}d}(y_2+y_2'),\quad q_a=e^{i\pi\tau_a},\quad q_b=e^{i\pi\tau_b}.
\end{eqnarray*}
\subsubsection{Example}
As an example let us check the obtained Green's function corresponding to the square impenetrable box  with the side $d$. If the initial state is \eqref{45} one can write
\begin{eqnarray}\label{32}\Psi(x_1,x_2,t)&=&\int\limits_{-d}^{d}\int\limits_{-d}^{d}A\sin(n_1y_1)\sin(n_2y_2)\frac{A^2}{4}\left(\theta_3(\frac{\pi}{2d}(x_1-y_1),q)-\theta_3(\frac{\pi}{2d}(x_1+y_1),q)\right)\nonumber\\
 &\cdot& \left(\theta_3(\frac{\pi}{2d}(x_2-y_2),q)-\theta_3(\frac{\pi}{2d}(x_2+y_2),q)\right)dy_1dy_2
 \end{eqnarray}
 \begin{eqnarray*}&=&\frac{A^3}{4}\int\limits_{-d}^{d}\sin(n_1y_1)\sum\limits_{N_1=1}^{\infty}q^{N_1^2}(\cos(\frac{N_1\pi}{d}(x_1-y_1))-\cos(\frac{N_1\pi}{d}(x_1+y_1)))dy_1\nonumber\\\nonumber
&\cdot& \int\limits_{-d}^{d}\sin(n_2y_2)\sum\limits_{N_2=1}^{\infty}q^{N_2^2}(\cos(\frac{N_1\pi}{d}(x_2-y_2))-\cos(\frac{N_1\pi}{d}(x_2+y_2)))dy_2
\end{eqnarray*}
The integrals are the following
\begin{eqnarray*}&&\int\limits_{-d}^{d}\sin\left(\frac{\pi N_1 }{d}y_1\right)(\cos(\frac{N_1\pi}{d}(x_1-y_1))-\cos(\frac{N_1\pi}{d}(x_1+y_1)))dy_1\\
&=&\frac{d}{\pi N_1}(2\pi N_1-\sin(2\pi N_1))\sin\left(\frac{\pi N_1 }{d}x_1\right).
\end{eqnarray*}
Hence, substituting the latter result in \eqref{32} we can write
\begin{eqnarray*}\Psi(x_1,x_2,t)&=&A^3d^2\sum\limits_{N_1,N_2=1}^{\infty}q^{N_1^2+N_2^2}\sin\left(\frac{\pi N_1 }{d}x_1\right)\sin\left(\frac{\pi N_2 }{d}x_2\right).
\end{eqnarray*}
Since $A=1/d$ we get the known state
\begin{eqnarray*}\Psi(x_1,x_2,t)=A\sum\limits_{N_1,N_2=1}^{\infty}e^{-\frac{i\pi^2 t}{2md^2}(N_1^2+N_2^2)}\sin\left(\frac{\pi N_1 }{d}x_1\right)\sin\left(\frac{\pi N_2 }{d}x_2\right).\end{eqnarray*}
 Hence, the obtained Green's function is correct.

\section{Summary}
In this last section we wish to summarize and point out the main results reached in this paper. Our first conclusion is that the separability of the Schr\"{o}dinger equation of an unconstrained system of two particles, coupled or not, generally brakes when boundary condition of geometric nature (holonomic constrains) are taken into consideration.
The main reason of such behavior may be traced back to the "coupling" get established between the "coordinates" as a consequence of the algebraic equations describing the same constraints.
We have illustrated these point writing the Schr\"{o}dinger equation of two noninteracting particles referred to $X_c$ and $x$, center of mass and relative motion coordinate, respectively.
By imposing generalized confinement conditions on the system, that is limiting the region of motion and assuming impenetrability conditions, one is lead to equations relating $X_c$ and $x$ which makes, impossible the separation of $X_c$ and $x$ in the problem. It is remarkable that searching the stationary state  of this constrained system exhibits the same mathematical formulation one should use to treat the problem of a single particle in a plane but confined in a domain whose shape depends on the boundary conditions imposed to the two-particle system.
Moreover, free particle motion on the plane is studied. The domain of the motion restricted by the impenetrable  walls  forming  the square, the rhombus, the triangle and the rectangle (billiards) are considered.  The billiards are  analyzed by studying the time evolution of the covariance of $X_c$ and $x$. To this end the Green's function expressed in terms of the Jacobi $\theta_3$ function is applied.

\end{document}